\documentclass{article}

\usepackage{cite} 
\usepackage{listings} 

\title{State Machine Model for The Update Framework (TUF)}
\author{
  Brian Romansky \and
  Thomas Mazzuchi \and
  Shahram Sarkani
}
\date{\today}

\begin{document}

\maketitle

\begin{abstract}
The Update Framework or TUF was developed to address several known weaknesses that have been observed in software update distribution and validation systems.  Unlike conventional secure software distribution methods where there may be a single digital signature applied to each update, TUF introduces four distinct roles each with one or more signing key, that must participate in the update process.  This approach increases the total size of each update package and increases the number of signatures that each client system must validate.  As system architects consider the transition to post-quantum algorithms, understanding the impact of new signature algorithms on a TUF deployment becomes a significant consideration.  In this work we introduce a state machine model that accounts for the cumulative impact of of signature algorithm selection when used with TUF for software updates.
\end{abstract}

\section{Introduction}
The Update Framework (TUF) is a method for distributing software updates to distributed client systems that was designed to specifically address many security issues and limitations that have resulted in threats against client systems \cite{SamuelJustin2010Skci}.  In prior work, we have shown that this method can be extended to meet the requirements of critical infrastructure systems \cite{RomanskyBrian2024ETUF}.  Adoption of the The Update Framework (TUF) as part of a software deployment system can have a significant impact on the overall network bandwidth and computational requirements for client systems \cite{LMS-v-XMSS}.  For distributed systems where network limitations or limited client system battery power may be important constraints, this can have a significant impact on the system design.  The choice of digital signature algorithm becomes even more impactful when post-quantum options are considered as the signature size and computational requirements may be significantly higher than they are for conventional digital signatures \cite{NSA-CNSA-2022r2}.

The interaction of the four roles in TUF, when combined with an expected series of updates required for a specific software package or device type, can result in a complex sequence of signature operations making it difficult to predict the number of signatures that will be required over the expected lifetime of a client system.  However, executing the sequence of steps for each of the TUF roles is relatively simple.  A state machine model can therefore be used to step through the operations to fully account for the cumulative impact of all digital signatures.

We introduce a model that performs this accounting.  The model is in the form of Python code that can implement any type of TUF deployment and account for the cumulative impact of a pre-planned sequence of timed update events.  The model accepts a list of digital signature algorithm parameters as input.  This allows for rapid evaluation of many different signature methods which can be compared when a consistent sequence of update events is executed.  

\section{Model Architecture}
The model is designed to execute the logic of TUF, but it does not actually perform the signature steps or produce the intermediate files for distribution.  The goal of the model is to perform comparisons among multiple different digital signature algorithm parameters.  Therefore, the actual size of the files being distributed is ignored, as this would be identical for all signature types.

The current implementation is focused on the impact of signature algorithm selection on the cumulative size of all files that must be distributed and downloaded by a client system as well as the computational cost for a single client to validate all signatures.  Centralized activity, such as key-pair generation and signature generation are not included, but they may be added in a future version.  The model further assumes that a client system chooses to download every signed update.  This assumption represents a ``worst case'' scenario.  In many real systems, client systems may only download and validate a subset of signed files generated by the TUF deployment system.  

Stateful Hash-Based Signatures (SHBS) are a class of post-quantum algorithms that have been specifically recommended for applications such as software update security \cite{800-208}.  The model accepts a maximum number of signatures as a parameter for each signature algorithm.  During execution, the state machine will keep track of the total number of signatures issued by each key pair and it will account for a ``rollover'' operation to replace keys that have reached their useful limit.  The rollover operation involves the logical replacement of the expired key with the Root roles signing and distributing a new file that endorses a new public key along with all other active public keys in the system.  The model reports on the total number of signatures for all roles and the total number of rollover operations that were encountered during an execution run.  

The model accepts three configurable parameters:
\begin{itemize}
  \item \textbf{TUF Architecture}:  In a typical TUF deployment, there is only one signing key for each role.  However, the TUF design allows for multiple signers for any role.  The model can account for TUF systems with any number of instances of each role, and each instance can have a unique digital signature algorithm.
  \item \textbf{Update Sequence}: The model takes in a list of update \emph{events}, defined as the date and time when a planned update occurs.  This list represents the times when new software must be distributed to all client systems.
  \item \textbf{Signature Algorithms}: The model accepts a list of digital signature algorithm types.  Each algorithm type consists of an algorithm name, public key size (in bytes), signature size (in bytes), maximum number of signatures, and validation effort (in millions of cycles).  
\end{itemize}

The intent of using a fixed sequence of events to define the update sequence is to allow for easy comparison across different architectural choices.  The model allows TUF system designers to test different options for the TUF role architecture and assign unique signature algorithms to each role.  The impact of these choices can be compared when a fixed sequence of update events is required by the software update process.  

The model has been developed in Python.  The appendix provides a source code listing with a brief description of each class and function.  Input parameter files are expected to be in Comma-Separated Value (CSV) format and minimal error checking is performed for input validation.  The output of the code is listed on standard output with a CSV structure which can be redirected to standard output.  The code can be configured to iterate through a list of different configuration options while using the same update event list.  When this is done, the output makes it easy to compare the impact of each configuration of TUF roles and signature algorithm choice on the performance requirements for a client system.

\section{Future Work}
This model was developed for a specific task of comparing the impact of many different SHBS configuration options when used with TUF for distributing software updates.  For this purpose, using a fixed sequence of update events was useful as it allowed for easy comparison across many different parameter choices.  In future work, it would be valuable to allow for the update event list to be driven by an integrated statistical model, such as a Poisson process, to allow for variations in the frequency of update events.  This would allow for the model to simulate many different variations and show the sensitivity of parameter choices to different update requirements.  

\appendix

\section{Class: SignatureAlgorithm}
\label{signature_algorithm}
The \texttt{SignatureAlgoritm} class describes a stateful digital signature algorithm with a specified set of parameters.  It has properties that define the size of a signature, size of a public key, computational cost for each verification, and the maximum number of signatures that can be supported by the algorithm.

A helper function is included that can read an array of signature algorithm definitions from a text file, structured as a Comma-Separated Value (CSV) file.  The file must be structured with each algorithm on a single line with column headers that define which data element is in the column.  The header text must match the exact text string defined in the class.  

\begin{lstlisting}[language=Python, basicstyle=\tiny]
import csv
from decimal import Decimal

class SignatureAlgorithm:
    def __init__(self, name, sig_size, pk_size, max_sigs, cost):
        self.name = name
        self.sig_size = sig_size
        self.pk_size = pk_size
        self.max_sigs = max_sigs
        self.cost = cost

    def __str__(self):
        return(self.name)

    # Reads a list of algorithms from a CSV file.  Assumes that the file
    # has header rows that exactly match the filed names listed here.
    # Returns the list of algorithms that were listed in the CSV file.
    @staticmethod
    def LoadFromCSV(file):
        algorithms = []
        with open(file, 'r') as csvfile:
            # This section will read-ahead the first 1024 bytes of the CSV
            # to capture the field names, then clean up the field names
            # before re-loading the file and using those field names.
            sniffer = csv.Sniffer()
            dialect = sniffer.sniff(csvfile.read(1024))
            csvfile.seek(0)
            reader = csv.DictReader(csvfile, dialect=dialect)
            fieldnames = [name.strip() for name in reader.fieldnames]
            reader = csv.DictReader(csvfile, fieldnames=fieldnames)
            
            #reader = csv.DictReader(csvfile)
            for row in reader:
                new_alg = \
                SignatureAlgorithm(row['Name'],
                                   int(row['Signature Size']),
                                   int(row['Public Key Size']),
                                   int(Decimal(row['Max Signatures'])),
                                   float(row['Computational Cost']))
                algorithms.append(new_alg)
        return(algorithms)

    # Finds a named algorithm in a list, returns the requested algorithm
    # (if it is present in the list)
    @staticmethod
    def FindInList(alg_name, algorithms):
        for alg in algorithms:
            if alg.name == alg_name:
                return(alg)
        raise ValueError("Requested algorithm type not found.")
\end{lstlisting}

\section{Class: TUFRole}
The \texttt{TUFrole} class is used to encapsulate the functionality of one of the TUF roles.  

\begin{lstlisting}[language=Python, basicstyle=\tiny]
class TUFrole:
    def __init__(self, name, role_type, algorithm):
        self.name = name            # identifier for the role
        self.role_type = role_type  # role type (Root, Timestamp, etc)
        self.algorithm = algorithm  # algorithm used for signatures
        self.num_sigs = 0           # initialize signature count
        self.reserve = False        # indicate that this role is not in reserve
        self.pending = True         # indicate that role requires an update
        self.rollover = True        # idicate rollover is required

    def __str__(self):
        return(f"{self.role_type} : {self.name} : sigs = {self.num_sigs}")

    def SignFile(self):
        self.num_sigs = self.num_sigs + 1

    def setReserve(self, state):
        self.reserve = state

    def getReserve(self):
        return(self.reserve)
\end{lstlisting}

\section{Class: TUFRepository}
The \texttt{TUFRepository} class describes an entire TUF repository.  It can contain multiple instances of each role.  The class allows for roles to be defined dynamically.  This means that during the execution of an update sequence, a new role may be added or an old one removed, provided that there remains at least one instance of each role.  This mechanism was used to model events where a SHBS key that had reached the end of its useful life and needed to be replaced.  

\begin{lstlisting}[language=Python, basicstyle=\tiny]
class TUFRepository:
    def __init__(self, name):
        self.name = name
        self.TUFroles = []
        self.accum_sig_size = 0
        self.accum_pk_size = 0
        self.accum_cost = 0.0
        self.accum_signatures = 0
        self.update_root = True   # indicates that a new root is needed

    def __str__(self):
        return(self.name)

    def showStats(self):
        print(f"Repo \'{self.name}\' has accumulated:")
        print(f"     {self.accum_sig_size} bytes of signatures")
        print(f"     {self.accum_pk_size} bytes of public keys")
        print(f"     {self.accum_cost} units of cost")
        print(f"     {self.accum_signatures} total signatures")

    def showStatsCSV(self):
        # Output is in columns in the following order:
        # <repo name>, <algorithm>, <sig bytes>, <pk bytes>, <net cost>,
        # <total sigs>
        print("%s, %d, %d, %d, %f, %d" %
              (self.name,
               self.accum_sig_size,
               self.accum_pk_size,
               (self.accum_sig_size + self.accum_pk_size),
               self.accum_cost,
               self.accum_signatures))

    def getAccumulatedBytes(self):
        return (self.accum_sig_size + self.accum_pk_size)

    def getAccumulatedCost(self):
        return (self.accum_cost)

    def addRole(self, name, role_type, alg):
        if (role_type == "Root" or
            role_type == "Timestamp" or
            role_type == "Snapshot" or
            role_type == "Target"):
            self.TUFroles.append(TUFrole(name, role_type, alg))
            self.update_root = True   # indicates that a root update is needed
            for r in self.TUFroles:
                if r.name == name and r.role_type == role_type:
                    r.rollover = True  # flag role to be added to a new root.txt
                    r.pending = True   # flag role as pending a update
                
    def listRoles(self):
        print(f"Repo \'{self.name}\' roles:")
        for r in self.TUFroles:
            print(f"   {r}")

    def removeRole(self, name):
        rval = 0
        for r in self.TUFroles:
            if r.name == name:
                self.TUFroles.remove(r)
                rval = rval + 1
        if rval > 0:
            self.update_root = True   # indicates that a root update is needed
        return(rval)

    # Check to see if any roles have run out of signatures
    # If they have, then stage them for a roll-over on the next timestamp
    def doRollover(self):
        n_rollovers = 0
        for r in self.TUFroles:
            # if a rule has reached max signatures and it is required to
            # perform a signature on the next timestamp, then stage it for
            # a rollover
            if (r.rollover == True or
                (r.num_sigs == r.algorithm.max_sigs and r.pending == True)):
                r.rollover = True
                r.num_sigs = 0
                n_rollovers = n_rollovers + 1
        return(n_rollovers)

    # Produce a timestamp.
    # Note that ALL roles that are pending a signature operation or rollover
    # are performed at this time.  
    def doTimestamp(self):
        # Perform a rollover - i.e. flag roles that need to be replaced
        if self.doRollover() > 0 or self.update_root:
            # If a rollover is required, then account for root signatures +
            # account for all public keys in the new root.txt file
            for r in self.TUFroles:
                self.accum_pk_size = self.accum_pk_size + r.algorithm.pk_size
                if r.role_type == "Root":
                    self.accum_sig_size = (self.accum_sig_size +
                                           r.algorithm.sig_size)
                    self.accum_cost = self.accum_cost + r.algorithm.cost
                    self.accum_signatures = self.accum_signatures + 1
                    r.num_sigs = r.num_sigs + 1
                r.rollover = False  # clear rollover flag (role is now updated)
                self.update_root = False # clear an update flag if it was set

        # Update the Targets.txt files for any targets that are updated
        n_updates = 0
        for r in self.TUFroles:
            if (r.role_type == "Target" and
                r.pending == True and
                r.reserve == False):
                # accumulate target signatures 
                self.accum_sig_size = self.accum_sig_size + r.algorithm.sig_size
                self.accum_cost = self.accum_cost + r.algorithm.cost
                self.accum_signatures = self.accum_signatures + 1
                r.num_sigs = r.num_sigs + 1
                r.pending = False
                n_updates = n_updates + 1

        # If any targets were updated, then also update Snaphot.txt
        if n_updates > 0:
            for r in self.TUFroles:
                if (r.role_type == "Snapshot" and
                    r.reserve == False):
                    self.accum_sig_size = (self.accum_sig_size +
                                           r.algorithm.sig_size)
                    self.accum_cost = (self.accum_cost +
                                       r.algorithm.cost)
                    self.accum_signatures = self.accum_signatures + 1
                    r.num_sigs = r.num_sigs + 1
                
        # For every timestamp role that is not flagged as a reserve role,
        # stage it for signing.
        n_stamps = 0
        for r in self.TUFroles:
            if r.role_type == "Timestamp" and r.reserve == False:
                self.accum_sig_size = (self.accum_sig_size +
                                       r.algorithm.sig_size)
                self.accum_cost = (self.accum_cost +
                                   r.algorithm.cost)
                self.accum_signatures = self.accum_signatures + 1
                r.num_sigs = r.num_sigs + 1
                n_stamps = n_stamps + 1

                
    # Stage an update for the target named "target"
    # Returns a count of the number of matching targets that were found
    def doUpdate(self, target):
        rval = 0
        #print(f"Doing an update on target \'{target}\'")
        # Search through all of the roles
        for r in self.TUFroles:
            # If a role name matches the target and it is a Target role
            if r.name == target and r.role_type == "Target":
                #print(f"Activating update on target \'{target}\'")
                r.pending = True
                rval = rval + 1

        # If you found a matching target, then also activate any Snapshots
        if rval > 0:
            for r in self.TUFroles:
                if r.role_type == "Snapshot":
                    r.pending = True

        return(rval)
\end{lstlisting}

\section{Wrapper Functions}
A collection of wrapper functions are used to set-up and execute a model for a single instance of a TUF repository over time.

\begin{itemize}
\item The function \texttt{date\_range()} produces a sequence of events.  By default there is a single event for every day that falls within the specified start and end date.  This sequence of events is used to trigger timestamps.  This function may be modified to create a different sequence of event types.  For example, timestamps could be changed to occur once per-week or once per-hour or per-minute.
\item The function \texttt{run\_sequence()} builds and executes a sequence of updates.  It accepts a CSV file name that lists the repeatable sequence of update events that will be used in the simulation.  The code shown here is configured to build one TUF repository with a single instance of each role.  It uses the same digital signature algorithm definition for each role.  Alternative versions of \texttt{run\_sequence()} have been used to model the assignment of unique digital signature parameters for each role.  
\item \texttt{main()} loads a list of algorithm definitions from a CSV file and then calls \texttt{run\_sequence()} for each algorithm type defined in the CSV file.  The result for each algorithm type is streamed to standard output and can be redirected to a file for analysis.  
\end{itemize}

\begin{lstlisting}[language=Python, basicstyle=\tiny]
from datetime import datetime, timedelta
from TUFroles import *
from TUFsequence import *

###############################################################################
# Function to create a sequence of days between two defined dates
def date_range(start_date, end_date):
    for n in range(int((end_date - start_date).days) + 1):
        yield start_date + timedelta(n)

################################################################################
# Function to build a repository and run a sequence of updates
def run_sequence(alg1, verbose, start, end, seq_file, dev_name):
    r = TUFRepository(dev_name)
    
    r.addRole("Root 1", "Root", alg1)
    r.addRole("Timestamp 1", "Timestamp", alg1)
    r.addRole("Snapshot 1", "Snapshot", alg1)
    r.addRole("Target 1", "Target", alg1)

    # Iterate through all dates in the range
    seq = EventList(start, end, seq_file)
    dates = seq.GetDates()
    events =  seq.GetEvents()

    for d in dates: # iterate through range of dates
        if d in events: # if this date matches an event
            if(verbose):
                print(f" - match {d.isoformat()}")
            r.doUpdate("Target 1") # require an update

        #for i in range(24):
        r.doTimestamp() # publish timestamp

    r.showStatsCSV()
    return()

###############################################################################
# Main function to run a model over a range of algorithms and dates
def main():
    alg_list = SignatureAlgorithm.LoadFromCSV('algorithms.csv')

    for alg in alg_list:
        print(f"{alg}, ", end='')
        run_sequence(alg, False, "2020-01-01", "2021-01-01",
                     "device_A.csv", "Device_A")
\end{lstlisting}

\bibliographystyle{plain} 
\bibliography{Romansky-TUF_model} 

\end{document}